\documentclass[10pt]{article}
\usepackage{amsmath}
\usepackage{amssymb}
\usepackage{graphicx}

\usepackage{amsfonts, amssymb, amsmath, amsthm, multicol}
\usepackage[colorlinks=true, pdfstartview=FitV, linkcolor=blue,
citecolor=blue, urlcolor=blue]{hyperref}
\usepackage[usenames]{color}
\definecolor{Red}{rgb}{1,0.05,0}
\definecolor{Grn}{rgb}{0.1,0.7,0.1}
\definecolor{Blu}{rgb}{0.1,0.1,0.6}
\definecolor{Org}{rgb}{1,0.45,0}
\definecolor{Vio}{rgb}{0.6578,0,0.9478}
\definecolor{Mag}{rgb}{1,0.2,0.3}

\usepackage{extsizes}
\usepackage[super,sort&compress,comma]{natbib} 
\usepackage[version=3]{mhchem}
\usepackage[left=1.25 in, right=1.25 in , top=1.5 in, bottom=1.5 in]{geometry}
\usepackage{fnpos}
\usepackage[english]{babel}
\usepackage{array}
\usepackage[usenames,dvipsnames]{xcolor}
\usepackage{setspace}
\usepackage{hyperref}
\usepackage{siunitx}
\usepackage{bm}
\usepackage{multirow}
\usepackage{float}
\usepackage{amsmath}
\usepackage{amssymb}

\newcommand{\footremember}[2]{%
	\footnote{#2}
	\newcounter{#1}
	\setcounter{#1}{\value{footnote}}%
}
\newcommand{\footrecall}[1]{%
	\footnotemark[\value{#1}]%
}

\usepackage{epstopdf}
\usepackage{footnote}
\definecolor{cream}{RGB}{222,217,201}

\author{%
	Zhao Pan\footremember{splash}{Splash Lab, Mechanical and Aerospace Engineering, Utah State University, Logan, UT, US 84322}\footremember{cofirst}{These authors contributed equally to this work.}\footremember{email}{To whom correspondence should be addressed: panzhao0417@gmail.com and/or taddtruscott@gmail.com}
	\and Floriane Weyer\footremember{grasp}{GRASP, CESAM,  Physics Department, University of Li\`ege, B-4000 Li\`ege, Belgium.}\footrecall{cofirst}%
	\and Williams G. Pitt\footremember{bill}{Department of Chemical Engineering, Brigham Young University, Provo, UT, US 84601.}
	\and Nicolas Vandewalle\footrecall{grasp} 
	\and Tadd T. Truscott\footrecall{splash} \footrecall{email}
	}

\title{Drop on a Bent Fibre \vspace{1 cm}}
\vspace{2 cm}
\date{}
\vspace{2 cm}

\begin{document}
\renewcommand{\abstractname}{\vspace{-\baselineskip}}
	\maketitle

	\vspace{1 cm}		 
	\begin{abstractname}
		Inspired by the huge droplets attached on cypress tree leaf tips after rain, we find that a bent fibre can hold significantly more water in the corner than a horizontally placed fibre (typically up to three times or more). The maximum volume of the liquid that can be trapped is remarkably affected by the bending angle of the fibre and surface tension of the liquid. We experimentally find the optimal included angle ($\sim \SI{36}{^\circ}$) that holds the most water. Analytical and semi-empirical models are developed to explain these {\color{black}counter-intuitive} experimental observations and predict the optimal angle. The data and models could be useful for designing microfluidic and fog harvesting devices.
	\end{abstractname}
	\vspace{1 cm}

	\section{Introduction}
	Droplets attached to fibres are common in daily life and can be found in many situations (e.g., droplets hang on and/or between thin pine needles \cite{duprat2012wetting}; a dog can remove the droplets caught in fur or hair by shaking violently \cite{dickerson2012wet}; droplets move along moss awns~\cite{pan2016upside} and cactus spines \cite{ju2012multi}). Perhaps the first to record the attach-detach behavior of droplets of critical size on plants was the famous Chinese poet Tu Fu who observed  \textit{``... Heavy dew beads and trickles, Stars suddenly there, sparse, next aren't ...}'' \cite{du1989selected}. The mesmerizing motion he penned 1260 years ago is of large dew droplets attached to plants (probably bamboo leaves based on the context, see supplemental materials) that fall off when the droplets are heavier than a critical mass. Recently, the physics of droplets attaching to fibres has been widely studied. Examples include, the geometric profiles of droplets on a fibre in a gravitational field~\cite{mei2013gravitational, wu2006droplet}, the dynamic wetting behavior on a fibre~\cite{seveno2004liquid}, the wetting of two crossed fibres~\cite{sauret2014wetting}, the wetting of two nearby flexible fibres~\cite{duprat2012wetting}, and the movement of droplets on inclined fibres~\cite{gilet2009digital,gilet2010droplets}.  A more thorough understanding of these common observations will have an impact on industrial applications (e.g., pharmaceutical production \cite{weyer2015compoundmanipulations}, and fog harvesting in arid regions \cite{pan2016upside, park2013optimal}).
	
\begin{figure}[tbh]
		\centering
		\includegraphics[width=.55\linewidth]{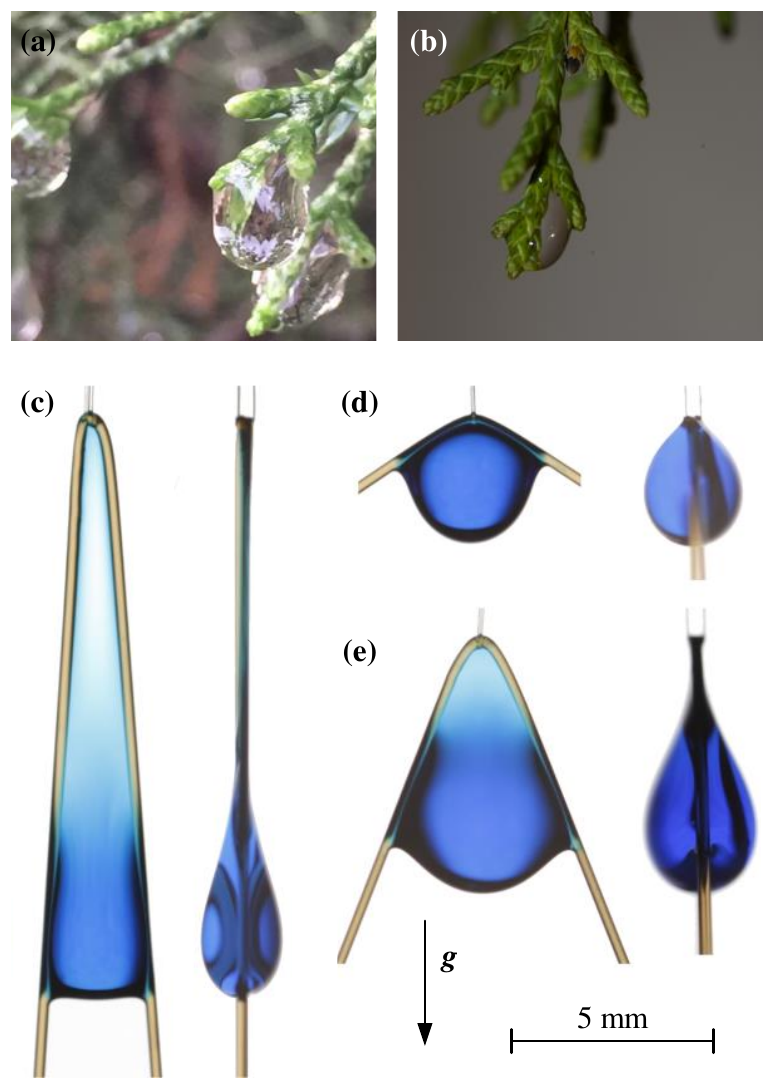}
		\caption{(a) Droplets attached on the leaves of a cypress tree after rain in Ditan Park, Beijing, China. (b) In the lab, a droplet is trapped on the interior fork of juniper leaf tips (collected from Cache Valley, UT) where the volume of the droplet is $\sim$ \SI{60}{\micro\liter}. 
			(c-e) Photographs of the morphology of droplets (SDS solution) on bent fibres (2b = \SI{250}{\micro \meter}) with various angles, {\color{black}front views on the left and side views on the right}: (c) $\theta = 4.25^{\circ}$; (d) $\theta = 58.2^{\circ}$; (e) $\theta = 21.5^{\circ}$.}
		\label{fig:morphology}
	\end{figure}

	Within the broad range of these investigations, Lorenceau et al.~\cite{lorenceau2004capturing} reported the maximum volume of a droplet on a horizontal fibre, which focused on the fundamental question: how much liquid can be held by a horizontal fibre in a gravitational field? Instead, we are inspired by the large water droplets attached to the ``armpit-like'' locations of tree leaves (Fig.~\ref{fig:morphology}(a)\&(b)); and seek to answer the question how much liquid can be held by a \emph{bent} fibre in a gravitational field? The volume of the droplets are about \SI{60}{\micro\liter}, which is significantly larger than can be supported by a simple horizontal fibre. This observation implies that by changing the geometry of a fibre (e.g., bending a fibre to create a corner), the critical volume of water can be significantly increased. Further, the previous theoretical work does not account for a bent fibre which is debatably more common in nature.	

	\section{Methods and materials}
	We used a circular rigid frame in which nylon fibres (diameters: \SI{160}{\micro\meter}, \SI{250}{\micro\meter} and \SI{350}{\micro\meter}) are stretched (SM Fig.1). The fibre was bent by first attaching it at two distinct points on the frame. Then, an $80$-\SI{}{\micro\meter}-diameter thin fibre was folded in half and attached vertically at the top of the frame, pulling the thicker fibre to a sharp corner. 
	The geometry and nomenclatures of a droplet-fibre system are shown in Table~\ref{fig:fig2}. 
	By modifying the position of the attachment points, the angle of the thick fibre was easily changed. In our experiments, we varied the half-angles, $\theta$, from $0^{\circ}$ to $90^{\circ}$. Prior to experiments, nylon fibres were cleaned with acetone and then with distilled water before drying in air. 
	
	{\color{black} Two liquid solutions (a solution of $0.01$ M sodium dodecyl sulfate (SDS) in water, a mixture of $25\%$ glycerol in water) and pure water were used in these experiments.} 
	In all solutions, dye was added for better visualization. The surface tensions of the liquids were measured using the pendant drop method via a CAM 200 goniometer (KSV Instrument Ltd). The surface tensions were $\gamma_{sw}= \SI{36.34}{mN/m}$ for the SDS mixture, $\gamma_{gw} = \SI{67.14}{mN/m}$ for the glycerol mixture, and {\color{black} $\gamma_{w} = \SI{71.97}{mN/m}$ for water}. Their densities and viscosities at $25^{\circ}\rm{C}$ were {\color{black}\SI{1.00e3}{kg/m^3}, \SI{1.19e3}{kg/m^3} and \SI{1.00e3}{kg/m^3}, and  \SI{0.89e-3}{Pa~s}, \SI{1.87e-3}{Pa~s},  and \SI{1.00e-3}{Pa~s} respectively}.  On a flat nylon surface the contact angles are $\sim$$40^{\circ}$ (SDS-water solution),  $\sim$$30^{\circ}$ (glycerol-water solution), {\color{black} and $\sim$$59^{\circ}$ (water). }
	
	Liquid was carefully measured by a micro-pipette and transported to the corner of the bent fibre via the thin vertical fibre (\SI{80}{\micro m}) which served as a drainage guide (SM Fig.~1). The capillary effect from the vertical thin fibre is negligible when the droplet is large (SM Fig.~2). The volume of the droplet was incrementally increased by \SI{1}{\micro\liter} until the droplet detached from the fibre (SM Fig.~3).
	
	
	\begin{table}[!tbhp]
		\centering
		\begin{tabular}{lll}
			\\
			\multirow{13}{*}{\includegraphics[width=0.25\linewidth]{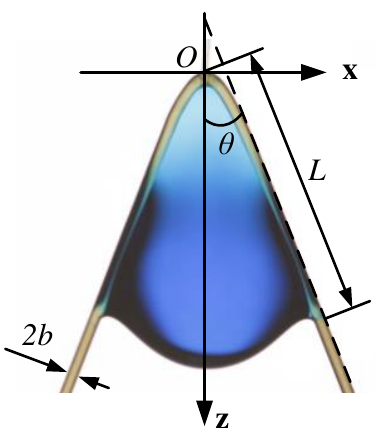}} & & \\
			& & \\
			&$2b$ & Thickness of the fibre \\
			& $\theta$ & Half angle between the fibres\\
			&$\Omega$ & Volume of the fluid \\
			&$L$ & Wetting length\\
			&$g$ & Gravity constant  \\
			&$\gamma$ & Surface tension of the fluid \\
			&$\lambda$ & Capillary length  \\
			&$Bo$ & Bond number  \\
			& & \\
			& & \\
		\end{tabular}
		\caption{ {\color{black} Image of a droplet on a bent fibre and a list of relevant nomenclature for this problem. Note that wetting length (L) is defined as the wetted length of one side of the fibre.}}
		\label{fig:fig2}
	\end{table}

	\section{Results}
	\subsection{Droplet morphology on a bent fibre} We hypothesize that the corner of a bent fibre can trap more liquid than a straight fibre, and the angle of the bent fibre affects the volume of the maximum liquid that the fibre can hold. 
	
	Fig.~\ref{fig:morphology}(c-e) shows side and front profiles of the maximum volume of an SDS-water solution that can be held by a fibre bent into three different angles. Fig.~\ref{fig:morphology}(c) shows a droplet attached to a sharp angled fibre ($\theta \approx 4.25^\circ$). The upper part of the droplet is mostly a thin film when viewed from the side (left of (c)), and is shaped like a triangle when viewed orthogonally (right of (c)). Fig.~\ref{fig:morphology}(d) presents a droplet trapped at the corner of a fibre with a large angle ($\theta \approx 58.2^\circ$) the droplet is nearly spherical, but stretched by the bent fibre. Fig.~\ref{fig:morphology}(e) shows a droplet attached to a bent fibre with an angle in between the other two ($\theta \approx 21.5^\circ$). We observe that as the angle decreases the bottom of the droplet becomes less spherical and that the top of the droplet thins, forming a film. Further, the droplet size of (e) is larger than in (c) and (d) where the only difference between all three cases is the angle of the bent fibre (more evidence shown in SM Fig. 3).
	
	
	\subsection{Experimental results}
\begin{figure}[tbh]
		\centering
		\includegraphics[width=0.7\linewidth]{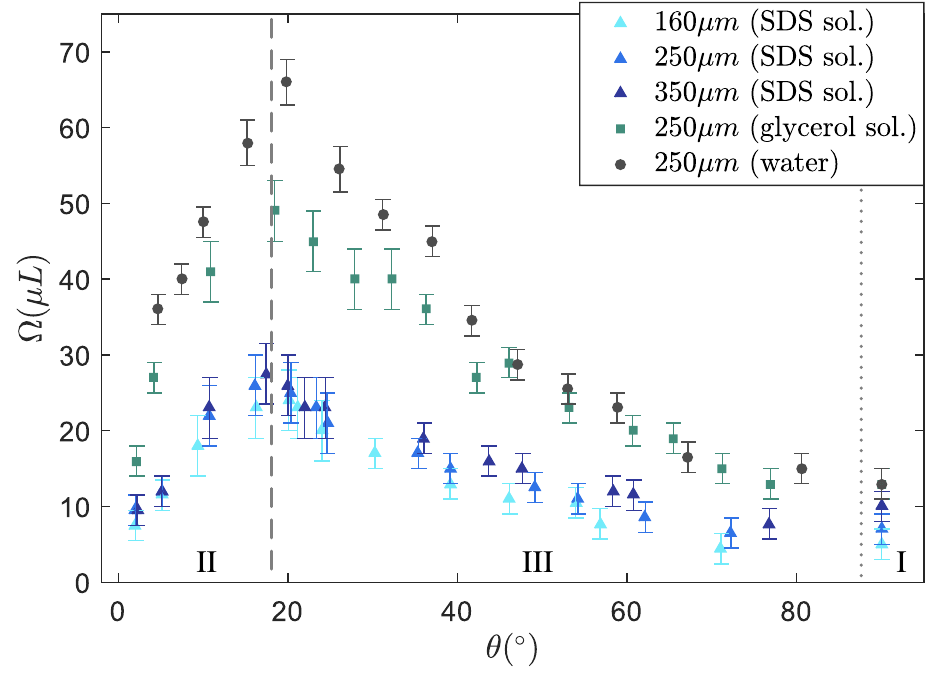}
		\caption{Raw data of the experiments. {\color{black}Data points} with error bars indicate the maximum volume of droplets that can be held on bent fibres with different angles $\theta$. {\color{black}Colors and shapes} represent different diameter wires and liquids as shown in the legend. The dashed line represents $\theta = 18^\circ$ where the largest volume can be held. The dotted line is a soft separation between regimes I and III. Uncertainty bands represent the total range of detachment and the markers represent the maximum measurable non-detachment. }
		\label{fig:rawdata}
	\end{figure}

	For each angle, we experimentally determined the maximum volume, $\Omega$, that can be held by the fibre (Fig.~\ref{fig:rawdata}) photographs of several data points are shown in SM Fig.~4. As expected, we observe that the fluid with the larger surface tension always has a larger maximum volume for any given $\theta$. {\color{black} For example the water and glycerol-water solution have higher surface tension values than the SDS solution, which correlates to higher values in Fig.~\ref{fig:rawdata}. }Irrespective of the fibre diameter (blue-hue triangles) {\color{black}or surface tension changes}, the same general trends are observed and we can split the results into three regimes. In regime I ($\theta \approx 90^\circ$), $\Omega$ tends to be independent of $\theta$. In regime II ($0^\circ < \theta \lesssim 18^\circ$), as $\theta$ increases $\Omega$ increases. In regime III ($18^\circ \lesssim \theta \lesssim  90^\circ$), as $\theta$ increases $\Omega$ decreases. Finally, the overall maximum volume of the liquid that can be held by any given fibre occurs at $\theta \approx 18^\circ$.

	\subsection{Modeling}
	\subsubsection*{Regime I}
\begin{figure}[!tbhp]
		\centering
		\includegraphics[width=0.45\linewidth]{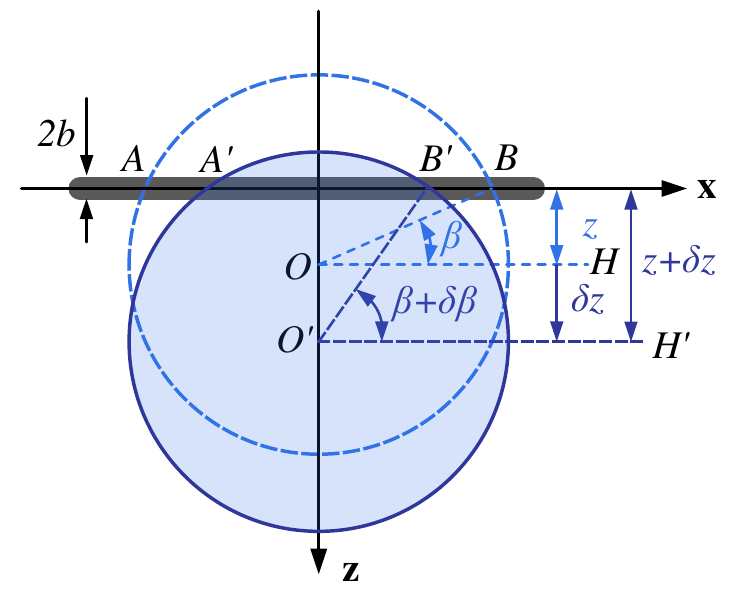}
		\caption{{\color{black}Geometry of a droplet on a fibre of diameter $2b$. The spherical droplet is originally positioned on a horizontal fibre ($\theta = \pi/2$) at equilibrium (represented by the dashed line with centre~$O$). The position of the droplet moves down to $\delta z$  when perturbed from equilibrium (represented by the filled circle with centre $O'$). $\beta$ is the angle between horizontal and the 3-phase point where the fibre exits the droplet ($\angle BOH$). The contact angle is not drawn.} 
		}
		\label{fig:perturbation1}
	\end{figure}

	We can develop mathematical models for the aforementioned observations, and start by considering the case of a droplet {\color{black} that wets and is} attached to a horizontal fibre (Fig.~\ref{fig:perturbation1}). The total free energy ($G$) attributed to the droplet on a horizontal fibre at equilibrium is
	\begin{equation}
	G = H_v \Omega+\gamma A_{LA} + \gamma_{SA} A_{SA} +\gamma_{LS} A_{LS} -\rho g \Omega z,  \label{eq:G0}
	\end{equation}
	where $H_v$ is the volumetric free energy of the fluid, $g$ is the gravitational constant, $\gamma$ is surface tension, $\rho$  is the density of the fluid, $\gamma_{SA}$ and $\gamma_{LS}$ are the interfacial energies at the solid-air interface and the liquid-solid interface, respectively. $A_{SA}$ and $A_{LS}$ are the areas of the solid-air interface and liquid-solid interface, respectively. $\Omega$ is the volume of the fluid, and $z$ is the center of the mass of the droplet. By assuming that Young's equation ($\gamma \cos \alpha + \gamma_{LS} =\gamma_{SA}$) is valid in this situation, \eqref{eq:G0} can be rewritten without the $\gamma_{LS}$ term:
	\begin{equation}
	G = H_v \Omega+\gamma A_{LA} + \gamma_{SA} A_{SA}+ (\gamma_{SA} - \gamma \cos \alpha )A_{LS} -\rho g\Omega z.
	\label{eq:G}
	\end{equation}
	When the {\color{black}position of droplet ($z$) is perturbed (from $z$ to $z+ \delta z$) by a vanishing distance $\delta z$ (herein, we use ``$\delta$'' as variational notation and see Fig.~\ref{fig:perturbation1} for geometric details)}, there is a change in $G$ {\color{black} (denoted as $\delta G$ hereafter)} which can be used to determine the stability criterion of the droplet-fibre system. 
	Noting that a positive perturbation $\delta z$ leads to a negative value of $\delta A_{LS}$ {\color{black}(meaning that the fiber is dewetted when the droplet moves downward)}, and invoking  $ \delta A_{SA} = -\delta A_{LS}$, {\color{black} variation of} \eqref{eq:G} leads to 
	\begin{equation}
	\delta G = H_v \delta \Omega + \gamma \delta A_{LA}  -  \gamma \cos \alpha \delta A_{LS} -\rho g\Omega \delta z.
	\label{eq:dG0}
	\end{equation}
	{\color{black}Noting that it maintains a constant volume of the droplet ($\delta \Omega =0$) after perturbation, \eqref{eq:dG0} becomes
		\begin{equation}
		\delta G = \gamma \delta A_{LA}  -  \gamma \cos \alpha \delta A_{LS} -\rho g\Omega \delta z, 
		\label{eq:dG3}
		\end{equation}
		where the first term on the right hand is the surface energy contribution from liquid-air interface change, the second term represents the contribution from the liquid solid interface change, and the third term is the gravitational energy change under perturbation.}
	
	
	The solid-liquid interface area is $A_{LS} \approx 4\pi bR\cos \beta$ {\color{black}(side surface area of a section of fiber ($\overline{AB}$, length $2R\cos\beta$), as shown in Fig.~\ref{fig:perturbation1})}, where $R$ is the radius of the droplet when assuming the droplet is spherical, {\color{black}and $\beta$ is the angle between horizontal and the 3-phase point where the fibre exits the droplet ($\angle BOH$, in Fig.~\ref{fig:perturbation1})}. Assuming that the shape of the droplet does not change after an infinitesimal perturbation of $\delta z$ ($\delta A_{LA} = 0$), the free energy change due to the perturbation can be derived from 
	\eqref{eq:dG3}: 
	\begin{equation}
	{\delta G}  \approx \gamma 4 \pi b \sin \beta {\delta \beta}  R \cos \alpha- \rho g \Omega{\delta z}.
	\end{equation}
	Noticing that {\color{black} $ R \cos \alpha \delta \beta / \delta z  \sim O(1)$}, we have the energy potential: 
	\begin{equation}
	\frac{\delta G} {\delta z} \approx \gamma 4 \pi b \sin \beta- \rho g \Omega.
	\label{eq:dGdz1}
	\end{equation}
	A critical condition approaches as $\delta G / \delta z \rightarrow 0$  and leads to a critical volume of the droplet:
	\begin{equation}
	\Omega \approx \frac{4 \pi \gamma b \sin \beta} {\rho g}.
	\label{eq:Omega1}
	\end{equation}
	Defining the capillary length of a fluid as $\lambda = \sqrt{\gamma / \rho g}$, \eqref{eq:Omega1} is identical to the equation  ($\sin \beta = \frac{1}{3} \frac{R^3}{b \lambda^2}$) found by Lorenceau et al.~\cite{lorenceau2004capturing}, which we also validated experimentally. {\color{black}Our method of obtaining \eqref{eq:dGdz1} \& \eqref{eq:Omega1} is outlined in more detail in the supplemental information. It is worth to note that, taking advantage of this energy based analysis, we are able to avoid the ``assumed equivalent'' configuration (two inclined fiber joining a droplet) used by \citet{lorenceau2004capturing}. Instead, we can directly analyze the stability of droplet held by a horizontal fiber. Nevertheless, the two modeling techniques (the energy based method employed in this paper and force balances used in \citet{lorenceau2004capturing} ) verify each other well.}
	
	A positive value of $\delta G / \delta z$ means that a droplet resists perturbation and tends to stay on the fibre with certain robustness (e.g., $\Omega$ is sufficiently small). At the critical condition ($\delta G / \delta z = 0$), the volume of the droplet is large enough that the contribution from the gravitational potential  (negative) tends to dominate over the contribution from interfacial energy (positive) given an infinitesimal perturbation. The destabilized droplet-fibre system then tends to fall off the fibre. On the other hand, when $\delta G / \delta z < 0$ (e.g., $\Omega$ is sufficiently large), a droplet falls off immediately {\color{black}due to the negative free energy potential}. 
	
	\eqref{eq:Omega1} implies that  the maximum possible volume of a liquid held by a horizontal fibre will occur when $\sin \beta$ approaches unity. 
	The maximum droplet size can now be estimated by normalizing \eqref{eq:Omega1} by a characteristic volume of a spherical droplet whose radius is the capillary length ($\tilde{\Omega} =\frac{4}{3} \pi \lambda^3$) yielding:  
	\begin{equation}
	\Omega^*_I = \frac{\Omega}{ \tilde{\Omega}} \approx 3\frac{b}{\lambda}, 
	\label{eq:OmegaStar1}
	\end{equation}  
	which we label as \textit{model I}. 
	The natural characteristic volume ($\tilde{\Omega}$) will also be used to normalize the other two models. 
	
	\subsubsection*{Regime II}
\begin{figure}[tbh]
		\centering
		\includegraphics[width=0.33\linewidth]{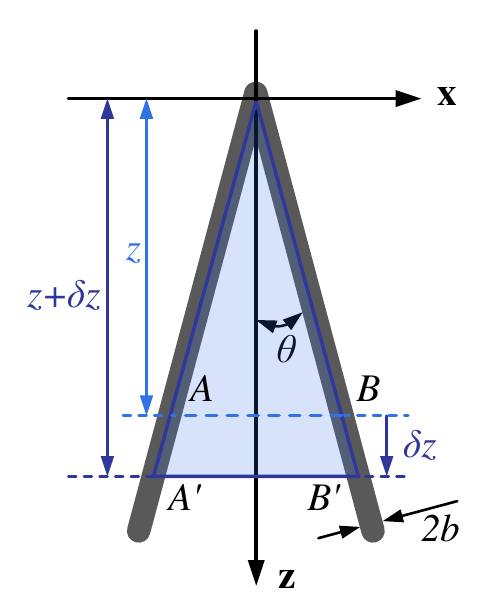}
		\caption{{\color{black}Geometry of a droplet between a bent fibre (diameter $2b$) with small $\theta$. The bottom of the droplet (solid line, $A'B'$) moved $\delta z$ downward after a perturbation from the original position (dashed line,~$AB$). The contact angle is not drawn. }
		}
		\label{fig:perturbation2}
	\end{figure}

	We continue our analysis with the same technique applied to the fibre bent at small angles (e.g., $ \theta \lesssim 18^\circ$). Note, we do not concern ourselves with the extreme case ($\theta = 0^\circ$, wetting of the parallel fibres) whose rich physics can be found in literature such as  ~\cite{protiere2013wetting}. At small angles, a droplet of critical size is characterized by a triangular thin film connected to the apex of the fibre (Fig.~\ref{fig:morphology}(c)). The area of the liquid-air interface is $A_{LA} \approx 2zL\sin \theta$, and the area of the liquid-solid interface is $A_{LS} \approx 4\pi b L$ (Fig.~\ref{fig:perturbation2}). Noting $z\approx L$ and $\sin \theta \approx \theta$ when $\theta$ is small, analysis of \eqref{eq:dG3} indicates that given an infinitesimal perturbation on the position of the droplet ($\delta z$) the free energy change of the droplet is 
	\begin{equation}
	\delta G \approx \gamma (4z\theta -4\pi b\cos \alpha)\delta z- \rho g \Omega \delta z.
	\label{eq:dG2}
	\end{equation}
	Our experiments show that in this regime, generally, the width of the bottom of the droplet {\color{black}($2z  tan \theta \approx 2 z\theta $)} is significantly larger than the diameter of the fibre ($2b$) (Fig.~\ref{fig:morphology}(c)). 
	
	In other words, the contribution to $\delta G$ from the fibre thickness is negligible compared to that of the liquid film between the fibre. 
	
	The critical state is approached as $\delta G/ \delta z \rightarrow 0$ and leads to a critical volume of the droplet at small angles: 
	\begin{equation}
	\Omega \approx  4\gamma L \theta / \rho g.
	\label{eq:Omega2}
	\end{equation}
	Again, when $\delta G / \delta z < 0$, the droplet total energy is reduced by perturbation and the droplet subsequently falls.  Normalizing \eqref{eq:Omega2} with characteristic volume ($\tilde{\Omega}$), we have created \textit{model II} that describes the critical volume for small angles:  
	\begin{equation}
	\Omega^*_{II} = \frac{\Omega}{\tilde{\Omega}} \approx  \frac{3L}{\pi \lambda} \theta =  \frac{3}{\pi} L_0 \theta, 
	\label{eq:OmegaStar2}
	\end{equation}
	where $L_0 = L/\lambda$ is  a length scale that characterizes the wetted length compared to the capillary length. 
	
	{\color{black} An alternative derivation of the model II based on force balance (similar to the method used in \cite{lorenceau2004capturing}) can be found in the supplemental information. It is worth noting that the free energy based derivation of model I and II itself allows access to more physical insights explicitly, which is not offered by force balance analysis. 
		
		Starting with one governing equation \eqref{eq:dG3}, we arrived at two models (model I, associated with \eqref{eq:Omega1} and model II, associated with \eqref{eq:Omega2}, respectively) by introducing two different assumptions. For example, \eqref{eq:Omega1} is derived by assuming that the droplet shape remains the same under perturbation; thus, the area of the liquid-air interface ($A_{LA}$) remains constant ($\delta A_{LA} \approx 0$, meaning that the \textit{first} term on the right hand side of \eqref{eq:dG3} is neglected, denoted as \textit{assumption I}). To arrive at \eqref{eq:Omega2}, we assumed that $\delta A_{LS} \approx 0 $ (\textit{assumption II}), meaning that the contribution from the change in solid-liquid interfacial area ($\delta A_{LS}$)  is negligible compared to the contribution from the liquid-air interface ($\delta A_{LA}$) when the droplet is perturbed. in other words, the \textit{second} term of the right hand side of \eqref{eq:dG3} vanishes. This mathematical symmetry provides explicit physical meaning for the models of regimes I and II. There is ``competition'' between the contritions of liquid-air interface and liquid-solid interface: when $\theta \rightarrow 0$, the liquid-air interface $A_{LA}$ is the dominant factor of the stability of the droplet-fiber system. However, for a droplet attached on a horizontal fiber ($\theta = \pi/2$), liquid-solid interface $A_{LS}$ dominates the physics. 
	}

	\subsubsection*{Empirical models}
	We now turn our attention to empirical results before formulating the model for regime III. {\color{black}The assumptions (\textit{assumptions} \textit{I} and \textit{II}) previously used are no longer valid for angles between $18^{\circ} \lesssim \theta \lesssim 90^{\circ}$.  
		Because $\delta A_{LA}/\delta z$ and $\delta A_{LS}/\delta z$ become complicated functions of the geometry of the fibre, the interfacial properties of the fluid, and more importantly, the geometry of the droplet itself (Fig.~\ref{fig:morphology}(e)).} The complexities make this minimum surface energy problem difficult to solve analytically; even approximate solutions such as \eqref{eq:Omega1} and \eqref{eq:Omega2} are not explicitly accessible. Therefore, we turn to experimental data to formulate a semi-analytical model.
	
\begin{figure}[!tbhp]
		\centering
		\includegraphics[width=0.65\linewidth]{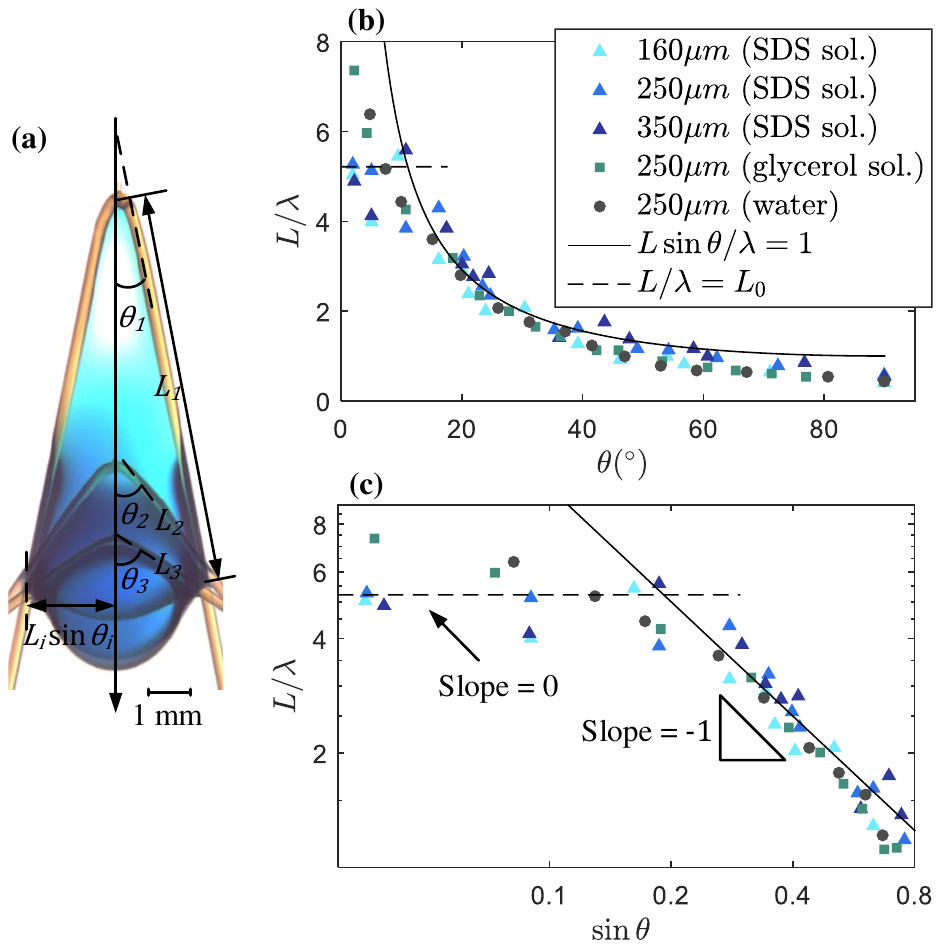}
		\caption{(a) Superimposition of photographs of {\color{black}SDS-water solution} droplets at critical state on fibre bent to various angles  ($\theta_i = 11.4^\circ, 36.8^\circ,~\&~ 67.8^\circ, i=1,2,3.$). 
			{\color{black}Non-superimposed photographs of these three droplets can be found in supplementary information.}
			(b) Normalized wetting length ($L/\lambda$) as a function of $\theta$. {\color{black}Triangle, square, or round} markers are experimental data, and the solid line represents the empirical model for $Bo =1$.  (c) $L/\lambda$ as a function of $\sin \theta$ in logarithm scale.}
		\label{fig:EmpiricalModel}
	\end{figure}

	We can illustrate that the characteristic length $\tilde{L} = L \sin \theta$ is approximately the same for a wide range of angles ($ 18^\circ \lesssim \theta \lesssim 90^\circ$) by superimposing photographs of droplets on a fibre as shown in Fig.~\ref{fig:EmpiricalModel}(a). Physically, $\tilde{L}$ is a length scale that characterizes the critical size of the droplet, and geometrically measures the half width of the droplet trapped between a bent fibre.  By comparing $\tilde{L}$ with the capillary length of the liquid we can formulate a normalized wetting length ($L^*$) as 
	\begin{equation}
	L^* = \tilde{L}  /\lambda = L \sin \theta/\lambda \approx 1.
	\label{eq:LStar}
	\end{equation}
	Experiments exhibit good agreement with the new parameter as shown in Fig.~\ref{fig:EmpiricalModel}(b\&c) (solid curve). The experimental data also reveal that the characteristic length of the drop-fibre system ($L \sin \theta$) is comparable to the  capillary length ($\lambda$) of the liquid. Not surprising, $\tilde{L}$ also yields a critical Bond Number of unity ($Bo=\rho g\tilde{L} ^2/\gamma=1$). In this context, the $Bo$ number can be considered a criteria of droplet stability. For example, $Bo>1$ indicates the weight of a droplet dominates and tends to detach from the fibre. A $Bo<1$ implies a stronger capillary force than gravity and thus a droplet stays on the fibre.

	Experimental data also show that when $\theta$ is small (e.g., $\theta \lesssim 10^\circ$), the normalized wetting length $L/\lambda$ does not strongly depend on $\theta$ (Fig.~\ref{fig:EmpiricalModel}(c)), which implies that $L/\lambda \propto \theta^0$. Data fitting (dashed line in Fig.~\ref{fig:EmpiricalModel}(b\&c)) reveals that $L_0 = L/\lambda \approx 5.1$, which provides a constant parameter for \textit{model II} (\eqref{eq:OmegaStar2}). 
	\begin{figure}[!tbhp]
		\centering
		\includegraphics[width=.5\linewidth]{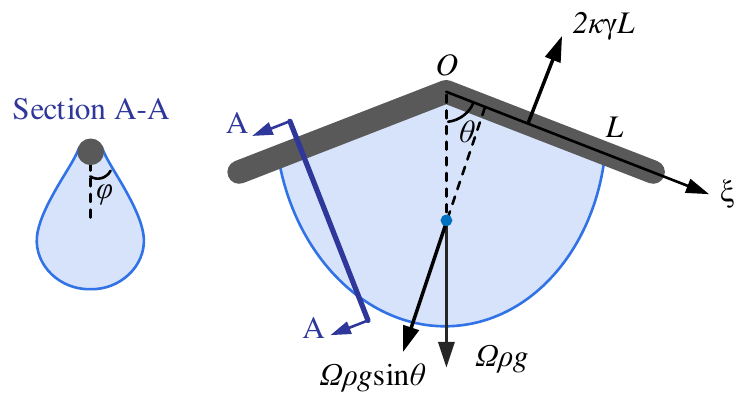}
		\caption{Free body diagram of the force balance on a droplet.  {\color{black} $L$ is the wetted length of one side of the fibre bent at an angle $\theta$ at the origin $O$. The capillary force is represented by $\gamma L$ and $\xi$ represents the axis of the fibre. A-A is a line indicating where the cross-section cutaway perpendicular to the fibre was made. The view of this cutaway is shown as Section A-A .} $\varphi$ is the local ``contact'' angle of the droplet.}
		\label{fig:freebody}
	\end{figure}
	
	\subsubsection*{Regime III}
	A natural way to model the critical state for angles $ 18^\circ \lesssim \theta \lesssim 90^\circ$ is to consider the force balance between the gravitational component and the surface tension provided by one side of the fibre (Fig.~\ref{fig:freebody}). The experimental observations reveal that at these angles a droplet slides down along one side of the fibre during detachment (Fig.~\ref{fig:detachmode}). Thus we can formulate the balance as 
	\begin{equation}
	\rho g \Omega \sin \theta = 2 \kappa \gamma L,
	\label{eq:model3}
	\end{equation}
	where $\kappa = \int_{0}^{L} \cos \varphi(\xi) \mathrm{d}\xi / L$  is a variable that measures the space-averaged effects of the local contact angle  ($\varphi(\xi)$) formed at the interface of the thick film of the droplet and the fibre as illustrated in the cross-section A-A of Fig.~\ref{fig:freebody}). 
	
	The droplet profile varies along the fibre as seen in Fig.~\ref{fig:morphology} ($\xi$-axis of Fig.~\ref{fig:freebody}), 
	Accordingly, $\varphi$ also varies along the fibre, and may even be a complicated function of $\theta$. Thus, $\kappa$ is difficult to calculate or measure, especially at critical states. However, $\kappa$ is not a practical or ``useful'' parameter even if it was a known parameter. Rather, there is no disadvantage to assuming that $\kappa$ is a constant which may lead to an acceptable and practical fitting parameter that is significantly less complex. 
	
	Substituting the empirical model of the wetting length \eqref{eq:LStar} into \eqref{eq:model3} leads to
	\begin{equation}
	\Omega \approx \kappa \frac{2 \lambda ^3}{\sin ^2 \theta}.
	\label{eq:Omega3}
	\end{equation}
	{\color{black}By choosing a simple $\kappa$ value $\kappa \approx 0.5$ we see decent agreement with the experiment especially when $\theta$ is large. } Normalizing the droplet volume by the characteristic volume ($\tilde{\Omega}$) leads to a semi-analytical model (\textit{model III}):
	\begin{equation}
	\Omega_{III}^* = \frac{\Omega}{\tilde{\Omega}} \approx  \frac{3 \kappa}{ 2 \pi \sin^2\theta}.
	\label{eq:OmegaStar3}
	\end{equation}
	
	\begin{figure}[tbh]
		\centering
		\includegraphics[width=.65\linewidth]{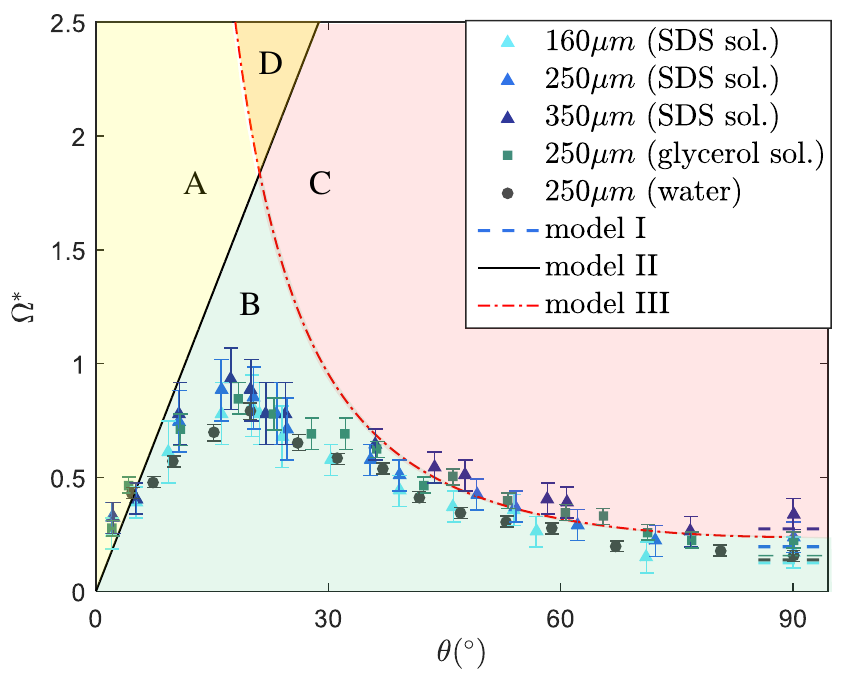}
		\caption{The non-dimensional maximum volume, $\Omega^*$, held by a bent fibre as a function of the half angle $\theta$ for several SDS solutions in blue and glycerol solutions in green (data from Fig.~\ref{fig:rawdata}). The three proposed models are compared to experimental data. Region B (green) indicates the conditions necessary for a droplet to remain on a bent fibre. Region A (yellow) indicates values where a droplet will fall off a bent fibre with small fibre angles (Fig.~\ref{fig:detachmode}(a)). Region C (pink) indicates values where a droplet stability first appears by sliding down one side of the fibre under small perturbations (Fig.~\ref{fig:detachmode}(b)). Region D indicates the transient regime between A and C. Maximum volume for experiments occurs near $\theta \approx 18^\circ$ and for theory $\theta \approx 21^\circ$. Note: model I is only applicable at $\theta = 90^\circ$ and dependent on fibre diameter as shown in the lower right corner.}
		\label{fig:OmegaStar}
	\end{figure}
	
	
	{\color{black}One heuristic way to think about the physical nature of regime III is to consider a spherical droplet with a horizontal fibre running through it. If the fibre is then bent inside the droplet with some small angle ($\theta<\pi/2$ in this case), then the wetted length of the fibre will become $2L/sin\theta > 2L$ and there will be a larger force attaching the droplet to the longer fibre, so the mass and the volume of the droplet can increase by $\sim1/sin\theta$. }
	
	\subsection{Model validation and discussion}
	In Fig.~\ref{fig:OmegaStar} we provide experimental validation for the three models. The data from four sets of experiments collapse onto a common trend when we utilize the non-dimensionalization of $\Omega^*$.  Model I (\eqref{eq:OmegaStar1}) shows good agreement to experiments with horizontal fibres ($\theta = 90^\circ$) as expected. {\color{black}Model III also shows good agreement with horizontal fibres ($\theta = 90^\circ$), which points to a subtle but important part of this modeling, that there is actually a smooth transition from Model I to Model III in the limit of $\theta = 90^\circ$  (see supplemental information for derivation).} Model II (\eqref{eq:OmegaStar2}) and model III (\eqref{eq:OmegaStar3}) show good agreement with the experiments for small and large angles, respectively. Based on model II and III, we can predict the optimal angle ($\theta_{opt}$) at which the maximum volume occurs. Comparing  \eqref{eq:OmegaStar2} with \eqref{eq:OmegaStar3} to eliminate $\Omega^*$ and noticing that $\sin \theta \approx \theta$ for relatively small angles, the optimal angle becomes
	\begin{equation}
	\theta_{opt} \approx \sin^{-1} \left( \frac{\kappa}{2 L_0}\right)^{1/3} \approx 21 ^\circ, 
	\end{equation}
	which is the theoretical prediction of the optimal angle. $\theta_{opt}$  maximizes the droplet volumes for fibres with arbitrary thickness and liquids with various surface tensions (intersection of models II \& III in Fig.~\ref{fig:OmegaStar}). Although the optimal angle prediction is sufficiently accurate, the experiments show that the realistic optimal angle is somewhat smaller ($\theta \approx 18^\circ$) than the model prediction. {\color{black} Additionally, the volumetric prediction over predicts by nearly a factor of 2.} This inconsistency could potentially be caused by i) the experiments may be sensitive to many other factors especially when the volume of the droplet approaches the maximum volume (e.g., local wettability of the fibre, vibration introduced by the pipette applicator, etc.); and/or ii) the models we developed may oversimplify the real physics (e.g., $\kappa$ is indeed affected by the droplet geometry). {\color{black}In our experiments, we have noticed both modes of droplet removal occurring at angles around 18$^\circ$. This is consistent with the asymptotic models developed herein. Fig. 6 illustrates this as $\theta$ approaches the critical angle either droplet detachment mode could occur.}

	\subsubsection{Droplet detachment behavior} 
\begin{figure}[tbh]
		\centering
		\includegraphics[width=0.65\linewidth]{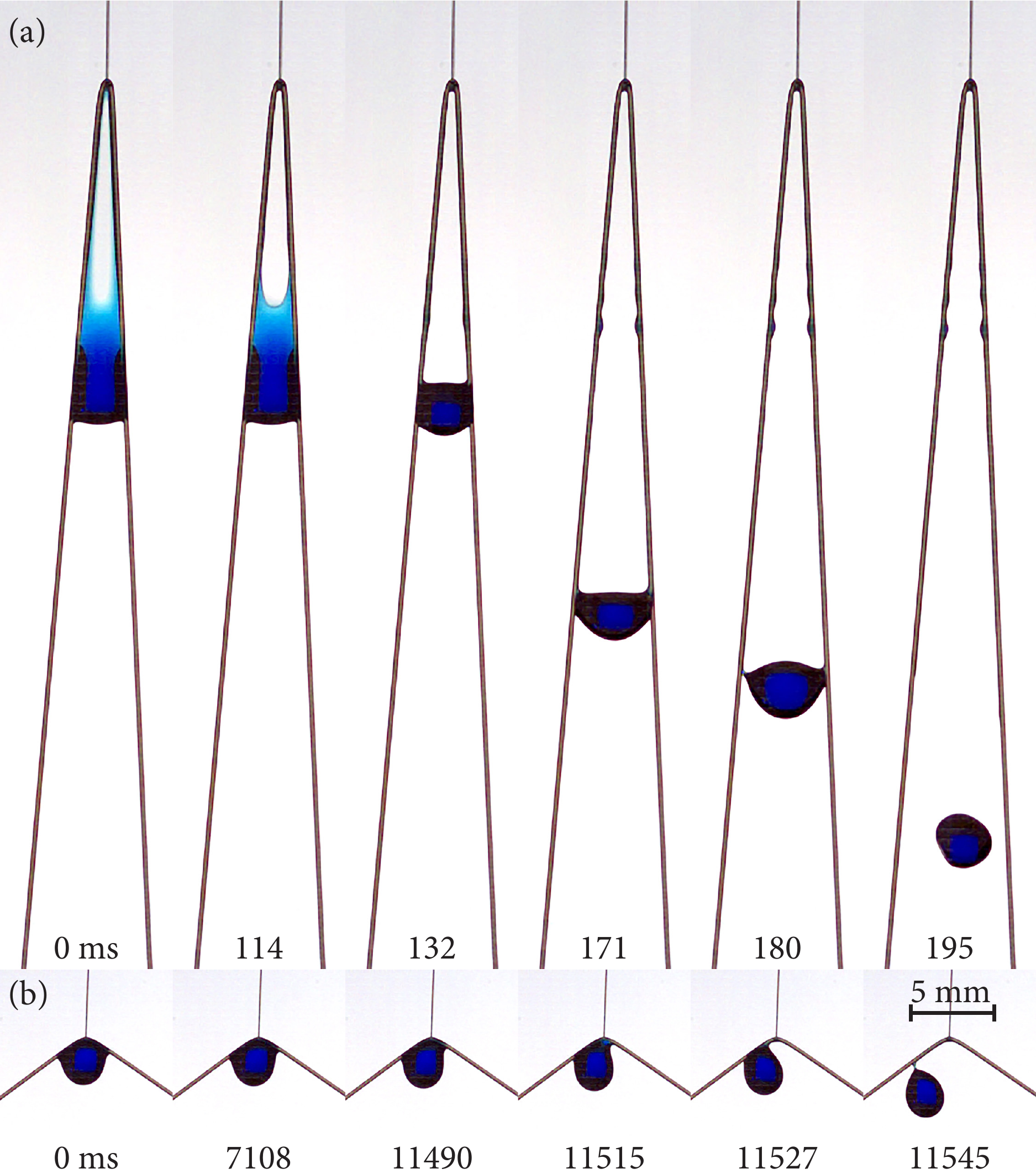}
		\caption{High speed photographs of the critical droplet volume on a bent fibre (Full videos found in SM video 1 \& 2). Time from the first droplet placement are labeled at the bottom of each image. The destabilized droplet detaches differently depending on the angle $\theta$. (a) The thin film on the top of the droplet breaks and the droplet falls symmetrically between the bent fibre, releasing from both sides nearly simultaneously ($\theta = 3.91^\circ$). (b) The droplet slides down along the left side of the fibre for larger $\theta$ angles ($\theta = 53.87^\circ$)}
		\label{fig:detachmode}
	\end{figure}

	More interestingly, model II and III split the $\theta$-$\Omega^*$ space into four regions. In region B (green region in Fig.~\ref{fig:OmegaStar}), droplets can be held steadily on a fibre. In regions A and C, droplets tend to detach from the fibre with two different modes. In region A (yellow area in Fig.~\ref{fig:OmegaStar}), when a droplet exceeds the critical volume predicted by model II, the thin film on the top of a droplet breaks and the droplet falls off the fibre  ($t =  114$ ms, Fig.~\ref{fig:detachmode}(a), supplemental video 1). This observation confirms one of the assumptions of model II that the main contribution of the droplet stability is the triangular film at the top of the droplet. In region C (red section in Fig.~\ref{fig:OmegaStar}), when a droplet is larger than the critical volume predicted by model III, the droplet slides down along one of the two sides of the fibre ($t = 11.49$ s, Fig.~\ref{fig:detachmode}(b), supplemental video 2). Region D (orange area in Fig.~\ref{fig:OmegaStar}) is the transition region where both modes could happen. 
	
	\section{Conclusion}	
	Inspired by the large droplets of water that can be held by bifurcated tree leaf tips after a rain storm (e.g., cypress trees) , we investigated the capability of a bent fibre to retain liquid droplets. Theoretical and semi-theoretical models were developed and validated experimentally. The volume of liquid droplets hanging on bent fibres is maximized when the half-angle $\theta \approx 18^\circ$. This volume is 3-fold more than can be suspended on a horizontal fibre. 
	
	\section*{Acknowledgements}
	Z.P. and T.T.T. thank the Utah State University Chemistry department for lending us a micro-pipette dispenser and showing us how to use it. Z.P. and T.T.T. thank N.V. and Dr. St\'ephane Dorbolo for securing funding for our visit to the University of Li\`ege. F.W. is financially supported by an FNRS grant.
	
	\section*{Author Contributions Statement}
	Z.P. and F.W. performed the experiments and drafted the first manuscript. Data were analyzed and the manuscript edited by all authors.
	
	\section*{Conflict of interest}
	The authors declare no conflict of interest.

	
	
	
	
	\bibliography{DropOnFiberLibrary} 

\providecommand*{\mcitethebibliography}{\thebibliography}
\csname @ifundefined\endcsname{endmcitethebibliography}
{\let\endmcitethebibliography\endthebibliography}{}
\begin{mcitethebibliography}{15}
\providecommand*{\natexlab}[1]{#1}
\providecommand*{\mciteSetBstSublistMode}[1]{}
\providecommand*{\mciteSetBstMaxWidthForm}[2]{}
\providecommand*{\mciteBstWouldAddEndPuncttrue}
  {\def\EndOfBibitem{\unskip.}}
\providecommand*{\mciteBstWouldAddEndPunctfalse}
  {\let\EndOfBibitem\relax}
\providecommand*{\mciteSetBstMidEndSepPunct}[3]{}
\providecommand*{\mciteSetBstSublistLabelBeginEnd}[3]{}
\providecommand*{\EndOfBibitem}{}
\mciteSetBstSublistMode{f}
\mciteSetBstMaxWidthForm{subitem}
{(\emph{\alph{mcitesubitemcount}})}
\mciteSetBstSublistLabelBeginEnd{\mcitemaxwidthsubitemform\space}
{\relax}{\relax}

\bibitem[Duprat \emph{et~al.}(2012)Duprat, Protiere, Beebe, and
  Stone]{duprat2012wetting}
C.~Duprat, S.~Protiere, A.~Beebe and H.~Stone, \emph{Nature}, 2012,
  \textbf{482}, 510--513\relax
\mciteBstWouldAddEndPuncttrue
\mciteSetBstMidEndSepPunct{\mcitedefaultmidpunct}
{\mcitedefaultendpunct}{\mcitedefaultseppunct}\relax
\EndOfBibitem
\bibitem[Dickerson \emph{et~al.}(2012)Dickerson, Mills, and
  Hu]{dickerson2012wet}
A.~K. Dickerson, Z.~G. Mills and D.~L. Hu, \emph{Journal of the Royal Society
  Interface}, 2012, \textbf{9}, 3208--3218\relax
\mciteBstWouldAddEndPuncttrue
\mciteSetBstMidEndSepPunct{\mcitedefaultmidpunct}
{\mcitedefaultendpunct}{\mcitedefaultseppunct}\relax
\EndOfBibitem
\bibitem[Pan \emph{et~al.}(2016)Pan, Pitt, Zhang, Wu, Tao, and
  Truscott]{pan2016upside}
Z.~Pan, W.~G. Pitt, Y.~Zhang, N.~Wu, Y.~Tao and T.~T. Truscott, \emph{Nature
  Plants}, 2016, \textbf{2}, 16076\relax
\mciteBstWouldAddEndPuncttrue
\mciteSetBstMidEndSepPunct{\mcitedefaultmidpunct}
{\mcitedefaultendpunct}{\mcitedefaultseppunct}\relax
\EndOfBibitem
\bibitem[Ju \emph{et~al.}(2012)Ju, Bai, Zheng, Zhao, Fang, and
  Jiang]{ju2012multi}
J.~Ju, H.~Bai, Y.~Zheng, T.~Zhao, R.~Fang and L.~Jiang, \emph{Nature
  communications}, 2012, \textbf{3}, 1247\relax
\mciteBstWouldAddEndPuncttrue
\mciteSetBstMidEndSepPunct{\mcitedefaultmidpunct}
{\mcitedefaultendpunct}{\mcitedefaultseppunct}\relax
\EndOfBibitem
\bibitem[Du and Hinton(1989)]{du1989selected}
F.~Du and D.~Hinton, \emph{The selected poems of Tu Fu}, New Directions
  Publishing, 1989\relax
\mciteBstWouldAddEndPuncttrue
\mciteSetBstMidEndSepPunct{\mcitedefaultmidpunct}
{\mcitedefaultendpunct}{\mcitedefaultseppunct}\relax
\EndOfBibitem
\bibitem[Mei \emph{et~al.}(2013)Mei, Fan, and Shou]{mei2013gravitational}
M.~Mei, J.~Fan and D.~Shou, \emph{Soft Matter}, 2013, \textbf{9},
  10324--10334\relax
\mciteBstWouldAddEndPuncttrue
\mciteSetBstMidEndSepPunct{\mcitedefaultmidpunct}
{\mcitedefaultendpunct}{\mcitedefaultseppunct}\relax
\EndOfBibitem
\bibitem[Wu and Dzenis(2006)]{wu2006droplet}
X.-F. Wu and Y.~A. Dzenis, \emph{Acta mechanica}, 2006, \textbf{185},
  215--225\relax
\mciteBstWouldAddEndPuncttrue
\mciteSetBstMidEndSepPunct{\mcitedefaultmidpunct}
{\mcitedefaultendpunct}{\mcitedefaultseppunct}\relax
\EndOfBibitem
\bibitem[Seveno \emph{et~al.}(2004)Seveno, Ogonowski, and
  De~Coninck]{seveno2004liquid}
D.~Seveno, G.~Ogonowski and J.~De~Coninck, \emph{Langmuir}, 2004, \textbf{20},
  8385--8390\relax
\mciteBstWouldAddEndPuncttrue
\mciteSetBstMidEndSepPunct{\mcitedefaultmidpunct}
{\mcitedefaultendpunct}{\mcitedefaultseppunct}\relax
\EndOfBibitem
\bibitem[Sauret \emph{et~al.}(2014)Sauret, Bick, Duprat, and
  Stone]{sauret2014wetting}
A.~Sauret, A.~D. Bick, C.~Duprat and H.~A. Stone, \emph{EPL (Europhysics
  Letters)}, 2014, \textbf{105}, 56006\relax
\mciteBstWouldAddEndPuncttrue
\mciteSetBstMidEndSepPunct{\mcitedefaultmidpunct}
{\mcitedefaultendpunct}{\mcitedefaultseppunct}\relax
\EndOfBibitem
\bibitem[Gilet \emph{et~al.}(2009)Gilet, Terwagne, and
  Vandewalle]{gilet2009digital}
T.~Gilet, D.~Terwagne and N.~Vandewalle, \emph{Applied Physics Letters}, 2009,
  \textbf{95}, 014106\relax
\mciteBstWouldAddEndPuncttrue
\mciteSetBstMidEndSepPunct{\mcitedefaultmidpunct}
{\mcitedefaultendpunct}{\mcitedefaultseppunct}\relax
\EndOfBibitem
\bibitem[Gilet \emph{et~al.}(2010)Gilet, Terwagne, and
  Vandewalle]{gilet2010droplets}
T.~Gilet, D.~Terwagne and N.~Vandewalle, \emph{The European Physical Journal
  E}, 2010, \textbf{31}, 253--262\relax
\mciteBstWouldAddEndPuncttrue
\mciteSetBstMidEndSepPunct{\mcitedefaultmidpunct}
{\mcitedefaultendpunct}{\mcitedefaultseppunct}\relax
\EndOfBibitem
\bibitem[Weyer \emph{et~al.}(2015)Weyer, Lismont, Dreesen, and
  Vandewalle]{weyer2015compoundmanipulations}
F.~Weyer, M.~Lismont, L.~Dreesen and N.~Vandewalle, \emph{Soft matter}, 2015,
  \textbf{11}, 7086--7091\relax
\mciteBstWouldAddEndPuncttrue
\mciteSetBstMidEndSepPunct{\mcitedefaultmidpunct}
{\mcitedefaultendpunct}{\mcitedefaultseppunct}\relax
\EndOfBibitem
\bibitem[Park \emph{et~al.}(2013)Park, Chhatre, Srinivasan, Cohen, and
  McKinley]{park2013optimal}
K.-C. Park, S.~S. Chhatre, S.~Srinivasan, R.~E. Cohen and G.~H. McKinley,
  \emph{Langmuir}, 2013, \textbf{29}, 13269--13277\relax
\mciteBstWouldAddEndPuncttrue
\mciteSetBstMidEndSepPunct{\mcitedefaultmidpunct}
{\mcitedefaultendpunct}{\mcitedefaultseppunct}\relax
\EndOfBibitem
\bibitem[Lorenceau \emph{et~al.}(2004)Lorenceau, Clanet, and
  Qu{\'e}r{\'e}]{lorenceau2004capturing}
{\'E}.~Lorenceau, C.~Clanet and D.~Qu{\'e}r{\'e}, \emph{Journal of colloid and
  interface science}, 2004, \textbf{279}, 192--197\relax
\mciteBstWouldAddEndPuncttrue
\mciteSetBstMidEndSepPunct{\mcitedefaultmidpunct}
{\mcitedefaultendpunct}{\mcitedefaultseppunct}\relax
\EndOfBibitem
\bibitem[Protiere \emph{et~al.}(2013)Protiere, Duprat, and
  Stone]{protiere2013wetting}
S.~Protiere, C.~Duprat and H.~Stone, \emph{Soft Matter}, 2013, \textbf{9},
  271--276\relax
\mciteBstWouldAddEndPuncttrue
\mciteSetBstMidEndSepPunct{\mcitedefaultmidpunct}
{\mcitedefaultendpunct}{\mcitedefaultseppunct}\relax
\EndOfBibitem
\end{mcitethebibliography}
	\bibliographystyle{rsc} 

\end{document}